\documentclass[aps,prl,reprint,superscriptaddress,longbibliography,nofootinbib]{revtex4-1}
\usepackage[english]{babel}
\usepackage{siunitx}  
\usepackage{graphicx} 
\usepackage{miller}   
\usepackage[percent]{overpic} 
\usepackage{pifont}
\usepackage[utf8]{inputenc}
\DeclareSIUnit\rydberg{Ry}

\begin{document}


\title{Structural modification of thin Bi\hkl(1 1 1) films by passivation and native oxide model}


\author{Christian \surname{König}}
\affiliation{Tyndall National Institute, University College Cork, Lee Maltings, Cork T12 R5CP, Ireland}
\author{Stephen \surname{Fahy}}
\affiliation{Tyndall National Institute, University College Cork, Lee Maltings, Cork T12 R5CP, Ireland}
\affiliation{Department of Physics, University College Cork, College Road, Cork T12 K8AF, Ireland}
\author{James C. \surname{Greer}}
\affiliation{Nottingham Ningbo New Materials Institute and Department of Electrical and Electronic Engineering, University of Nottingham Ningbo China, 199 Taikang East Road, Ningbo, 315100, China}

\date{\today}

\begin{abstract}
The structure of thin terminated Bi\hkl(1 1 1) films of approximately \SI{1}{\nano\meter} thickness is investigated from first principles.
Our density functional theory calculations show that covalent bonds to the surface can change the orientation of the films completely.
For thicker films, the effect is limited to the surface only.
Based on these observations, we further present a simple model structure for the native oxide and chemically similar oxides, which form a protective capping layer, leaving the orientation of the films unchanged.
The advantages of this energetically favorable layered termination are discussed in the context of the films' technological exploitation in nanoelectronic devices.
\end{abstract}


\maketitle

\section{Introduction}
The semimetal bismuth is a material with many unusual electronic properties and as such has been studied intensively for many years.
A small overlap of the valence band
with the conduction band
by only a few \si{\milli\electronvolt} leads to a very low density of states and charge carrier density at the Fermi level \cite{hofmann2006}.
Furthermore, the effective mass of electrons moving along the $C_3$ symmetry axis in the bulk material can be exceptionally small and leads to a large de Broglie wavelength for the free carriers of several tens of nanometer \cite{hofmann2006}.
Therefore bismuth is an ideal prototype material for studies on quantum confinement in thin films or nanowires.
Confinement effects are expected to enable a new generation of nanoscale electronic devices \cite{gity2017,gity2018} by opening a bandgap in the electronic structure and thus making the material effectively semiconducting.
Semiconductor-like behavior was confirmed by measurements of the temperature dependent conductivity, e.g.,~in Refs.~\cite{fei2010,xiao2012,luekermann2013,aitani2014,zhu2016,kroeger2018}.
The expected quantum confinement relies on the assumption that the crystal structure in the nanoscale system does not change considerably compared to the bulk and that there are no conducting surface states.
These effects could both prevent the transition into the semiconducting phase.

The rhombohedral crystal structure of bulk bismuth consists of stacked bilayers parallel to the \hkl(1 1 1) plane.
Within a bilayer, each bismuth atom has a strong bond to three surrounding atoms.
The bonds between adjacent bilayers are comparatively weak and thus the bilayers also define the natural cleavage plane of the crystal.
The result is a threefold rotationally symmetric hexagonal surface.
A detailed discussion of the crystal structure and the Bi\hkl(1 1 1) surface is given in Refs.~\cite{hofmann2006,jona1967,moenig2005}.
In this work, we focus on the structural properties of thin bismuth films grown on a Si\hkl(1 1 1)-$7\times7$ substrate.
It is well known that the preferred growth direction of these films is \hkl(1 1 1) as observed in, e.g., Refs.~\cite{nagao2000,gity2017}.

During the deposition of the first few layers of material, various peculiar phenomena can be observed.
A wetting layer can be found between the Si\hkl(1 1 1)-$7\times7$ substrate and the bismuth film.
The interaction between wetting layer and film is found to be weak and the wetting layer is considered to be chemically inactive; for details we refer to Refs.~\cite{shioda1993,nagao2000,kim2001,saito2004}.
Moreover, very thin films of the order of less than ten bilayers of material show either a different orientation or an allotropic phase \cite{sadowski2003,nagao2004,kammler2005,nagao2005,hirahara2006,sadowski2006,yaginuma2007,oh2008,koroteev2008,bian2009,mccarthy2010,kowalczyk2011,kowalczyk2012,xiao2012,kowalczyk2014,kowalczyk2017}.
In summary, the \hkl(110) orientation or related structures can be found for many different substrates. 
Specifically on Si\hkl(1 1 1)-$7\times7$, a very similar allotropic phase which resembles the structure of black phosphorus is observed.
Alternative paired layer structures for films on highly oriented pyrolytic graphite are discussed in the literature as well \cite{kowalczyk2012}.
The allotropic films have a thickness of around \SI{1}{\nano\meter} which is in the technologically relevant regime \cite{gity2017}.
The deposition of more material leads to a reorientation towards the hexagonal Bi\hkl(111) surface.

In the following, we specifically discuss the effect of passivation and oxidation on the structure of the films.
From a device perspective, the formation of a native oxide could ideally be exploited as a passivation layer as it is common practice in silicon technology.
We show that for thin bismuth films this issue is closely related to a change in the crystal structure.
Note that although the properties of thin bismuth films and the corresponding surfaces are studied intensively in the literature, often these con\-si\-de\-ra\-tions are restricted to vacuum conditions and do not take into account any interaction with the environment.
Likewise, the formation of the native oxide is not very well studied and different stoichiometries are reported in the literature \cite{benbow1976,puckrin1990,leontie2002}.
So far, the atomistic structure is unclear and the oxide might also be amorphous.
However, recent measurements for a thin Bi\hkl(111) film suggest that the native oxide's stoichiometry is $\text{Bi}_{2}\text{O}_{3}$ \cite{gity2018}.
As the structure of thin films depends on the substrate, it is an intriguing question if or how the structure of the film changes at the interface to the oxide and what the oxide itself could look like.
In this paper we will concentrate on the structure of the films and postpone an investigation of the consequences for the passivation of surface states to future work.
Note that in the context of applications, e.g.,~in nanoelectronic devices, the surface is inevitably exposed to atmosphere or layers of dielectric material.

\section{Computational details}
We use density functional theory (DFT) to determine the properties of thin bismuth films within the framework of the generalized gradient approximation (GGA) and a Perdew-Burke-Ernzerhof (PBE) functional as implemented in the QuantumATK software package \cite{perdew1996,atk2018.06,soler2002}.
The electron density is represented by a lo\-ca\-lized basis set.\footnote{High-accuracy version as described in Ref.~\cite{smidstrup2017}.}
To include the effect of spin-orbit coupling, fully-relativistic pseudopotentials \cite{hamann2013,schlipf2015} are employed.
These treat the outer $15$ electrons of each Bi atom explicitly as valence electrons.
The thin Bi\hkl(1 1 1) films have a he\-xa\-go\-nal surface symmetry and thus are represented as slabs in a hexagonal cell for which a $\Gamma$-centered $6 \times 6 \times 1$ k-point grid was employed.
The density mesh cutoff for all films containing oxygen was set to \SI{350}{\rydberg}, while for all other systems, \SI{150}{\rydberg} was found to be sufficient to converge the total electronic energy and interatomic forces.
The atomic positions were optimized with an accuracy of $\pm\SI{0.05}{\electronvolt\per\angstrom}$ in the remaining forces.
To avoid any interaction between the periodic images of the film, the height $c$ of the simulation cell was adjusted so that the vacuum perpendicular to the surface was approximately \SI{20}{\angstrom} for all calculations.
A similar cell without vacuum was used to obtain the bulk lattice parameters.

The bilayered bulk crystal structure is well reproduced by DFT ($6 \times 6 \times 2$ k-points in the hexagonal cell)
although the weak interaction between the bilayers allows for some variation of the interlayer distance \cite{cantele2017}.
We obtain an equilibrium lattice parameter $a_{\text{Bi}} = \SI{4.532}{\angstrom}$ which is in very good agreement with the literature \cite{jona1967,hofmann2006,moenig2005}.
We note however that differences in the energy eigenvalues corresponding to the single particle electronic states are less accurate and that predictions made by DFT for the electronic band structure -- especially regarding the metallic, semimetallic or semiconducting character -- are in general not reliable.
Nevertheless, the direct and indirect band gap obtained by our GGA calculations is in agreement with previous studies \cite{aguilera2015} which show an improvement compared to the local density approximation.
We investigate the properties of thin films with a thickness of three or more bismuth bilayers, i.e.,~with a minimum thickness of approximately \SI{1}{\nano\meter}.
In this regime, a large enough band gap for the technological exploitation in nanoelectronic devices can be anticipated \cite{gity2017,gity2018}.
Furthermore, for the thickness under consideration, experimental data \cite{nagao2004,kammler2005,nagao2005,yaginuma2007}
suggests that the lattice mismatch between silicon and bismuth leads to a compressive strain of $1.3\%$ with respect to bulk Bi in the plane of the film.
We therefore set the lattice parameter to $a = \SI{4.473}{\angstrom}$.
Note that even then the Bi and Si surfaces do not match perfectly and periodicity with respect to the substrate is achieved only for a $6 \times 6$ supercell.
Therefore, instead of modeling the substrate on which the film grows explicitly, we only constrain the lattice parameter and terminate both sides of the free-standing bismuth slabs in order to passivate dangling bonds.
Thus we also avoid the formation of a wetting layer.
We note that the unterminated thin Bi\hkl(1 1 1) films with the given lattice parameter do not transform into an allotropic phase automatically by relaxation.
Furthermore, experimental results \cite{gity2017} indicate that the films may be polycrystalline.

\section{Hydrogen passivation}
\begin{figure}
\includegraphics{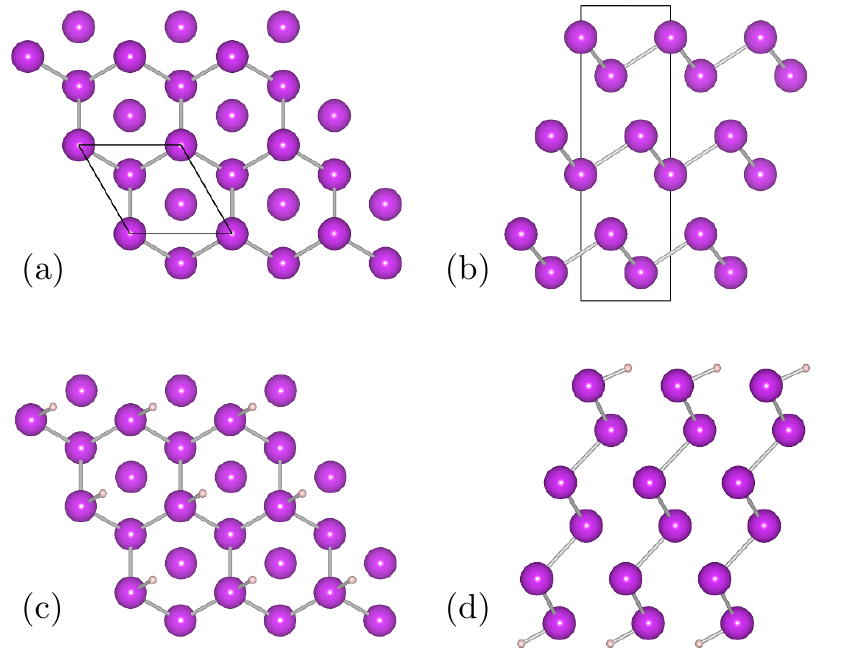}
\caption{ \label{fig:film}
(a) Hexagonal bulk bismuth, top view corresponding to the Bi\hkl(1 1 1) surface.
Only the bonds in the first bilayer are shown.
The borders of a single cell are drawn in black.
(b) Hexagonal bulk, side view. Bonds in and between bilayers can be clearly distinguished by their respective length.
Only the intralayer bonds are shown in the picture.
Note that the unterminated thin films have the same structure, even if strained to $a=\SI{4.473}{\angstrom}$.
(c) H-terminated Bi\hkl(1 1 1) surface before relaxation.
Hydrogen was attached so that the bond has approximately the orientation of an interlayer bond.
(d) Terminated film after relaxation, side view.
A reorganization of the structure changes the orientation of the bilayers to Bi\hkl(1 0 0).}
\end{figure}

Figures \ref{fig:film}(a) and \ref{fig:film}(b) show the hexagonal bulk cell of bismuth.
Assuming the equilibrium lattice parameter, the intralayer bonds $l$ are $\SI[separate-uncertainty=true]{3.10(1)}{\angstrom}$ long and thus can be clearly distinguished from the interlayer bonds which have a length of $l' = \SI[separate-uncertainty=true]{3.58(1)}{\angstrom}$ and are not drawn in the image.
These latter weak bonds are broken at the surface of the Bi\hkl(1 1 1) film.
We investigate the use of hydrogen to terminate the surface as it is common practice for, e.g., semiconductor surfaces.
In general, this reduces the computational cost considerably compared to modeling a complicated oxide layer.
H also is the chemically least complex monovalent termination possible which makes it suitable to study its chemical interaction with the surface of the thin films.

In Fig.~\ref{fig:film}(c) we show how we terminate the three-bilayer thick film with hydrogen atoms.
To start with, the surface atoms are terminated with one single hydrogen atom each on both sides of the film.
Therefore the area density of H on the surface is similar to the spacing of the surface atoms on the silicon substrate.
The bond direction approximately coincides with one of the interlayer bonds.
We find after relaxation at the constrained lattice parameter $a=\SI{4.473}{\angstrom}$ in Fig.~\ref{fig:film}(d) that the interaction of the Bi\hkl(1 1 1) surface with hydrogen is significant as it changes the structure of the thin film considerably.
In contrast to the unterminated case, the angle between the bilayers and the film's surface is increased from \SI{0}{\degree} -- the in-plane position -- to approximately \SI{70}{\degree}, i.e.,~almost perpendicular to the plane.
This corresponds to a reorientation of the bilayers from Bi\hkl(1 1 1) to Bi\hkl(1 0 0).
Note that the hydrogen atoms terminate the edges of each bilayer and form intra- instead of interlayer bonds, i.e.,~they substitute for the next bismuth atoms within the new bilayers.
The intralayer bonds drawn in Fig.~\ref{fig:film}(d) are in the range between \SI{3.12}{} and \SI{3.15}{\angstrom}.
The interlayer bonds $l'$ vary slightly more with one bond of approximately \SI{3.37}{\angstrom} and two bonds of approximately \SI{3.56}{\angstrom}.
Nevertheless, it can be clearly distinguished between the bilayers which reveals a reorientation throughout the thin film which is induced by the interaction with hydrogen at the surface.

After the reorientation the interlayer bonds are shorter, suggesting that the bilayers are closer to each other than in the bulk material.
The observation of two different bond lengths also implies that the slab is sheared: if the bilayer in Fig.~\ref{fig:film}(a) is moved in-plane with respect to the other bilayers (unbonded atoms in the picture), then the distances to the three nearest neighbors in the adjacent bilayer are no longer equal.
It is reasonable to assume that this shear in the film is due to the constraints of the crystal cell geometry.
Note that, as the interlayer bonds are weak, it is not surprising that they are more likely to deform than the intralayer bonds.
Depending on the initial position of the H atoms on the surface, i.e.~their position before relaxation, local minima in the total free energy can be found which are energetically separated from the global minimum by about \SI{450}{\milli\electronvolt} per cell (containing six Bi and two H atoms).
Figure \ref{fig:film}(d) shows that the bonds on the top and bottom surface of the optimized structure are antiparallel with respect to each other.
However, if the initial geometry is far from this ideal position, the system relaxes into a frustrated state.
This is plausible as the structural change is driven by the H bonds and the reorganization therefore starts from both surfaces and propagates through the film.
If the orientation of the newly formed bilayers do not match, they form a disordered structure where they meet in the middle.

\begin{figure}
\includegraphics{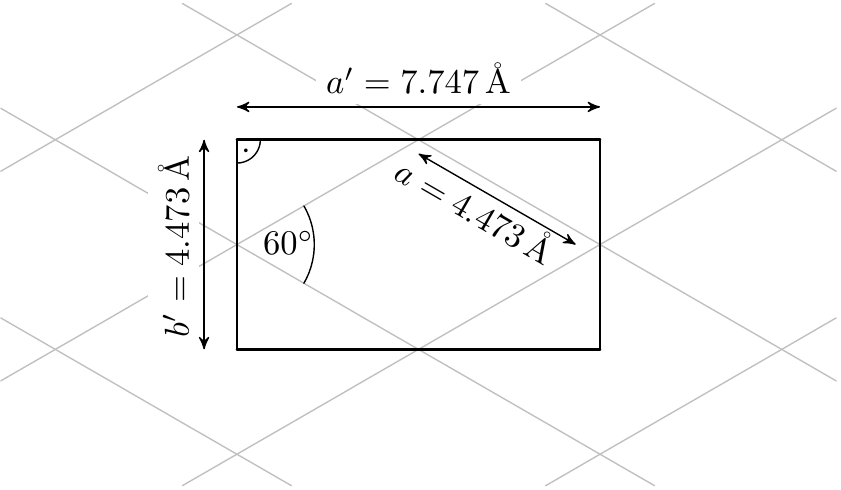}
\caption{ \label{fig:supercell}
Orthorhombic supercell (black) on the hexagonal surface lattice of a Bi\hkl(1 1 1) film (drawn in gray).
By reference to Fig.~\ref{fig:film}(d), we have seen that the orthorhombic cell is more appropriate regarding the relaxed structure of the film and imposes less constraints on the positions of the atoms.
Note that this cell is still consistent with the $1.3\%$ compressive strain due to the Si\hkl(1 1 1) substrate.
}
\end{figure}

\begin{table}[b]
\caption{ \label{tab:EperAtom}
Total free energy per atom in different crystal cells.}
\begin{ruledtabular}
\begin{tabular}{lcc}
                & hexagonal cell                       & orthorhombic cell \\
\hline\\
$a$ constrained & $0$ (reference)                      & $\SI{- 2}{\milli\electronvolt}$/atom \\
$a$ optimized   & $\SI{-17}{\milli\electronvolt}$/atom & $\SI{-46}{\milli\electronvolt}$/atom \\
\end{tabular}
\end{ruledtabular}
\end{table}
It is already apparent from Fig.~\ref{fig:film}(d) that the rotational $C_3$ symmetry of the surface is lost.
Instead, the edges of the bilayers are visible as stripes on the surface and an orthorhombic (super-) cell similar to Fig.~\ref{fig:supercell} seems to be a better choice.
Assuming an appropriate starting condition, i.e.,~orientation of the H-bonds, the bilayers can align to the sides of the cell and artificial constraints by the hexagonal geometry of the unterminated surface are removed during the relaxation which allows for a reduced stress and lower energy of the optimized structure.
Nevertheless, we find that the total free energy per atom can only be reduced by \SI{2}{\milli\electronvolt} in the orthorhombic cell, see Table \ref{tab:EperAtom}.
Note that the effect is more pronounced if the lattice parameter is optimized and that the interlayer bonds in the orthorhombic cell with optimized $a$ are almost equal which corresponds to reduced shear.

\begin{figure*}
\includegraphics{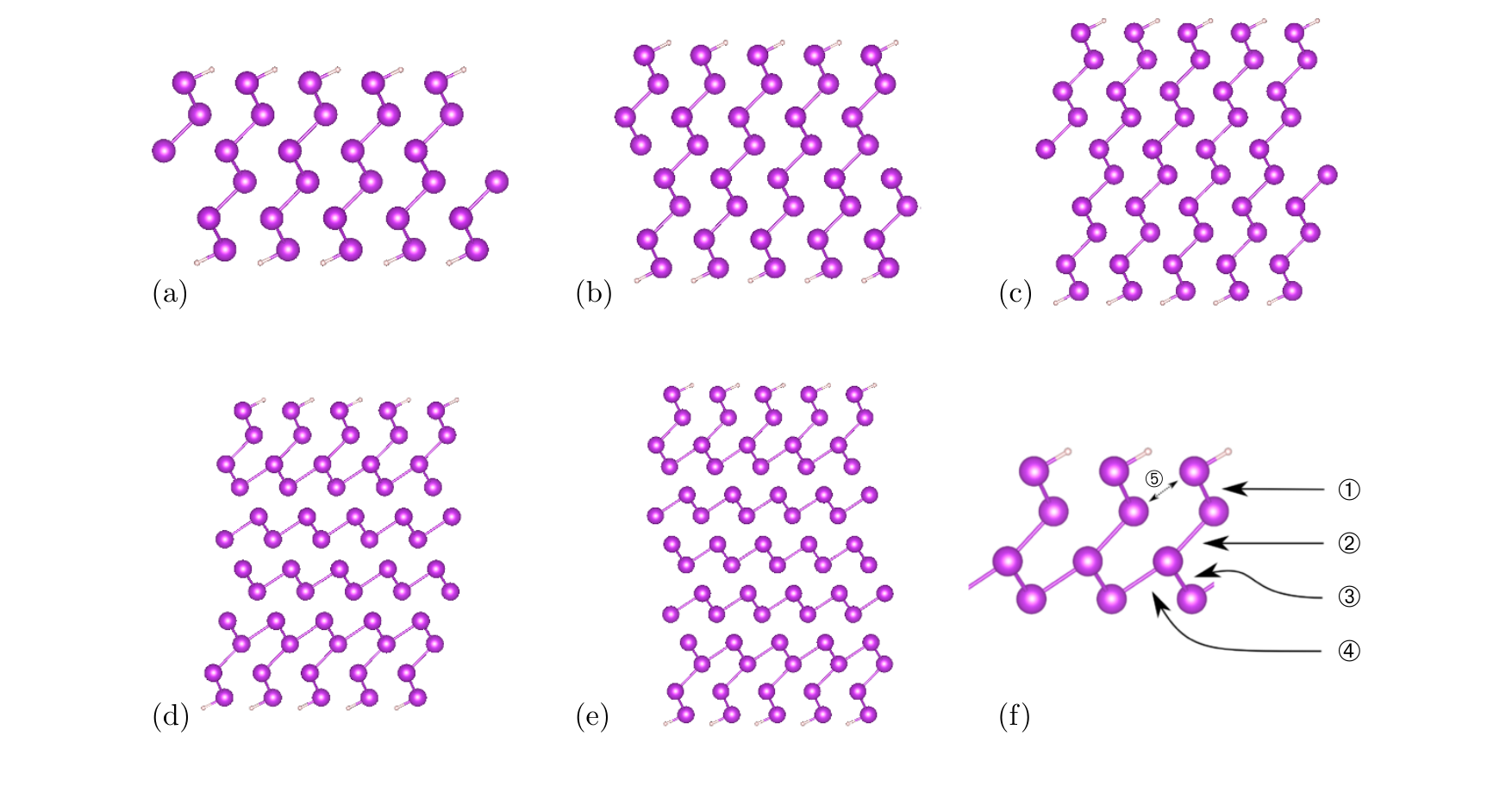}
\caption{ \label{fig:thickness}
Thin films with hydrogen termination for $3$ up to $7$ bilayers are shown in (a)--(e).
The orientation in the middle of the film changes back to \hkl(1 1 1) if the film is more than five bilayers thick.
(f) shows the outer layers of these structures which are clearly affected by the termination and where the overall structure does not depend much on the thickness of the film.}
\end{figure*}

\begin{figure*}
\includegraphics{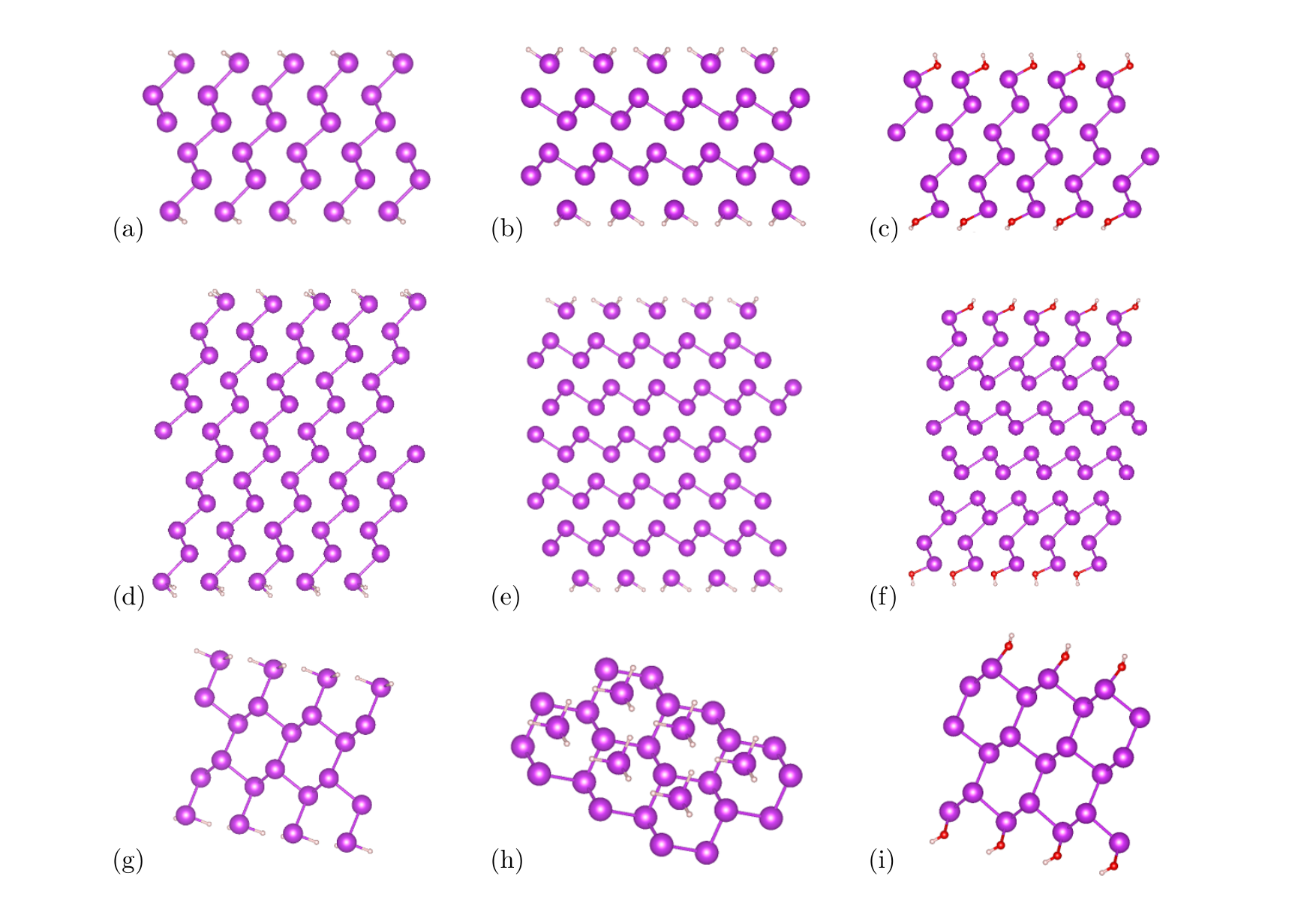}
\caption{ \label{fig:XH_XOH_3-6bl}
Thin films with alternative surface terminations. (a), (b), and (c) show the three-bilayer thick films terminated with two H atoms, three H atoms and one -OH molecule per surface atom, respectively. Only the termination with three H deviates from the previous results significantly because it keeps the bilayer orientation intact. (d), (e), and (f) show the corresponding six-bilayer thick films.
Here the termination with two hydrogens induces a full reorientation while three H atoms keep the Bi\hkl(1 1 1) orientation.
The middle bilayers of the OH-terminated structure have a Bi\hkl(1 1 1) orientation, just like in the case of a single hydrogen where the transition occured between five and six bilayers.
Finally, (g), (h), and (i) show a single bilayer of the above three bilayer films including the termination to elucidate the bonding scheme within the structures. This corresponds to a side view or -- in case of the three H termination which has a different orientation -- a top view of the films.}
\end{figure*}

Given the experimental evidence that thin bismuth films on Si\hkl(1 1 1)-$7\times7$ eventually grow in the \hkl(1 1 1) direction, a transition back to this \hkl(1 1 1) orientation can be expected with increasing film thickness -- similar to the allotropic phases discussed in the introduction.
To address the question when this transition occurs, we relaxed films with a thickness between 3 to 12 bilayers.
We use the orthorhombic supercell in Fig.~\ref{fig:supercell} with a $6\times10\times1$ k-point grid.
In an experimental situation, the lattice parameter is not necessarily constant everywhere in the film but asymptotically approaches the bulk value for bilayers far from the substrate.
This behavior is not compatible with the periodic boundary conditions of the calculations but it is reasonable to assume a constant $a$ for thin films with a thickness of only a few nanometers
(a six bilayer film with hydrogen termination is only \SI{2.5}{\nano\meter} thick).
Figure \ref{fig:thickness} shows the relaxed structures with $3$, $4$, $5$, $6$, and $7$ bilayers.
The cutoff for the maximum bond length to be displayed in the pictures was
increased
until all atoms had at least three bonds.
We find that the five-bilayer structure is still very similar to the three-bilayer structure.
Only when the thickness is increased to six bilayers, two \hkl(1 1 1) bilayers are seen to form in the middle of the film.
The adjacent bilayers on top and below these middle layers form a transition region while the structure close to each of the surfaces resembles the previous films.
If the film thickness is increased further, an equivalent number of \hkl(1 1 1) bilayers form in the middle of the film while the structure of the surface layers is preserved.
This also holds true for a $12$-bilayer thick structure (not shown here).
We note that in principle both bilayer orientations should be energetically equivalent in the middle of the film and the reorientation should therefore be complete.
However, a comparison of the bulk energy of both orientations at the given lattice parameters reveals that the Bi\hkl(1 1 1) orientation is energetically favorable. Thus the substrate induced strain eventually stabilizes the Bi\hkl(1 1 1) orientation.
This discussion does not take into account possible energy barriers which may also occur due to the reorganization in the interior of the film.
If the lattice parameters are not constrained, $a'$ in Fig.~\ref{fig:supercell} increases significantly and the reorientation can be observed throughout each of the films regardless of their thickness.

To obtain a numerical measure of the reorganization of the film's structure, it is worth examining the bond lengths in more detail.
We discuss the bonds in the middle of the films first because this is where the structural transition happens before we further investigate the surface layers.
Note that due to the finite precision of the method, small deviations in the bond lengths can be expected throughout the film after relaxation.
The intralayer bonds in the middle of the three-bilayer thick structure (orthorhombic cell) measure \SI[separate-uncertainty=true]{3.14(1)}{\angstrom}.
Although they are almost equal, two of these three bonds are slightly shorter than the remaining one.
Moreover, we find one interlayer bond with $l'=\SI{3.36}{\angstrom}$ and two with a length of $l'=\SI{3.58}{\angstrom}$, all measured relative to the position of the same atom.
For comparison, in the four-bilayer structure the two short intralayer bonds stay the same but the length of the longer one increases to \SI{3.22}{\angstrom}.
The interlayer bonds stay approximately the same with $l'=\SI{3.38}{\angstrom}$ and two times $l'=\SI{3.56}{\angstrom}$.
For the five-bilayer thick film, we find two intralayer bonds with $l=\SI{3.11}{\angstrom}$ and one with $l=\SI{3.27}{\angstrom}$ which is close to the shortest interlayer bond with $l'=\SI{3.36}{\angstrom}$. The other two interlayer bonds do not change and remain $l'=\SI{3.56}{\angstrom}$.
These observations show that with increasing thickness of the film one intralayer bond $l$ becomes longer while one $l'$ becomes shorter until eventually $l$ becomes $l'$ and vice versa.
This transition happens between five and six bilayers where for the first time two middle bilayers with \hkl(1 1 1) orientation can be observed.
The continuous transition of one intra- to an interlayer bond can also be clearly observed for the thicker films where the bonds in the middle of the film slowly converge towards a set of three equal intralayer bonds and three equally long interlayer bonds; the relevant values are given in Table \ref{tab:middleLayer}.

Secondly, we want to discuss the outer layers of the films with five or more bilayers thickness which all appear to have a similar structure.
The bond lengths for these structures are given in Table \ref{tab:bondLengths}.
Apparently, the length of each of these bonds converges quickly to a constant value.
Note that bond $3$ is considerably shorter than bond $4$ (by approximately \SI{5}{\percent}), i.e.,~the respective layer is not equivalent to a bulk bilayer.
The bonds equi\-va\-lent to bond $4$ but in the bilayers between the discussed top layers and the middle are shorter.
Their length is converging towards the value of bond $3$ with increasing thickness and towards the middle of the film.

\begin{table}
\caption{ \label{tab:middleLayer}
Bond lengths of an atom in the middle Bi\hkl(1 1 1) bilayers of the thick films
and the middle of the five-bilayer film, given in angstroms.
The bond lengths slowly converge towards two sets of three equally long intra- and interlayer bonds.
}
\begin{ruledtabular}
\begin{tabular}{ccccc}
                & 5 bilayers           & 6 bilayers           & 7 bilayers           & 12 bilayers \\
\hline\\
intralayer ($2\times$) & $3.11$ & $3.10$ & $3.09$ & $3.09$ \\
intralayer ($1\times$) & $3.36$ & $3.28$ & $3.19$ & $3.10$ \\
interlayer ($2\times$) & $3.56$ & $3.58$ & $3.57$ & $3.57$ \\
interlayer ($1\times$) & $3.27$ & $3.39$ & $3.47$ & $3.56$ \\
\end{tabular}
\end{ruledtabular}
\caption{ \label{tab:bondLengths}
Bond lengths in angstroms for the top layers of the thick films as shown in Fig.~\ref{fig:thickness}(f) for $5$, $6$, $7$, and $12$ bilayers.}
\begin{ruledtabular}
\begin{tabular}{ccccc}
                & 5 bilayers           & 6 bilayers           & 7 bilayers           & 12 bilayers \\
\hline\\
bond $1$ & $3.11$ & $3.11$ & $3.11$ & $3.11$ \\
bond $2$ & $3.20$ & $3.24$ & $3.25$ & $3.26$ \\
bond $3$ & $3.11$ & $3.11$ & $3.11$ & $3.10$ \\
bond $4$ & $3.36$ & $3.32$ & $3.29$ & $3.27$ \\
bond $5$ & $3.52$ & $3.56$ & $3.55$ & $3.55$ \\
\end{tabular}
\end{ruledtabular}
\end{table}

It is intriguing to investigate what happens if the surface density of hydrogen on the surface is increased.
We find that two hydrogen atoms per surface Bi can lead to the same type of reorganized structure for very thin films as the termination with a single H, see Fig.~\ref{fig:XH_XOH_3-6bl}.
The two strong bonds to hydrogen lead to a full reorientation of the six-bilayer thick film.
A totally different behavior is observed for the three hydrogen case where the orientation is always Bi\hkl(1 1 1).
The terminated atom is removed from its bilayer to form a $\text{Bi}\text{H}_3$ molecule which sits on the surface.
The remaining bismuth atoms in the film regroup -- two half bilayers from the top and bottom surfaces result in one additional bilayer -- and the bilayers in the interior of the film indeed have the same Bi\hkl(1 1 1) orientation as the unterminated film.

Hydroxyl termination can be employed alternatively to hydrogen and has the advantage that it is chemically similar in kind to the native oxide.
Interestingly, for a single -OH per surface atom, the same general trend towards a reorientation in the film as for a single hydrogen can be observed due to the termination, see also Fig.~\ref{fig:XH_XOH_3-6bl}.
The bilayers in a three-bilayer thick film change their orientation completely.
For six bilayers, only the top and bottom two bilayers feature the same surface layer structure which was already observed in the H case. 
The single bond to the monovalent hydroxyl molecule breaks the intralayer bonds in the top bilayers.
We note that mo\-le\-cu\-lar terminations like OH-groups introduce more degrees of freedom regarding how they can arrange on the surface compared to single atoms, especially when their density on the surface is high.
Thus the situation can become more complicated, for example, if they bond to neighboring passivants or lie flat on the surface, thus also interacting with nearby bilayers.
However, the tendency to change the bilayer orientation is the same.

It remains the question if this reorganization of the film can be observed in a real system. The DFT calculations show that bonding of single H atoms or OH groups are per se stable but do not take into account the formation of H$_2$ or H$_2$O. Following the line of thought in Refs.~\cite{reuter2001,reuter2003} we show in the supplemental material \cite{supplPRM1} that the H- and OH-terminated surfaces are likely not stable under typical experimental conditions. Film growth on the other hand inherently is a nonequilibrium process. From a general point of view we have shown that different terminations -- H and OH as well as others which are not shown here -- induce the same general behavior with regards to the structure as a function of the number of surface bonds. Therefore, if such covalent bonds to the film's surface can be realized, e.g.~at the interface between the film and a Si\hkl(111) substrate, the structure close to this surface changes which implies altered electronic properties.
We furthermore point out that the stability calculations for the OH-termination suggest that the oxide does not covalently bond to the film.
This observation is equally important because the passivation is often applied in order to remove conducting surface states by the formation of covalent bonds. In that regard, the efficacy of any form of stable and realistic passivation
is questionable and contradictory results can be found in the literature, see also Ref.~\cite{zhu2016}.
Thus, a simple termination with H as it is employed in the case of silicon or other semiconductors appears not to be appropriate and has to be used with care in DFT calculations due to its effect on the structure.

\section{Surface Oxidation}
A clean surface can only be maintained under experimental conditions in vacuum and not under ambient conditions where the sample is inevitably exposed to oxygen and water in the surrounding air.
Therefore, in order to avoid undesirable chemical reactions which can change the properties of a nanoelectronic device, the surface always has to be protected by a capping layer or a native oxide.
To the best of our knowledge, there is no existing model structure for the native oxide on a Bi\hkl(1 1 1) surface.
We will base our considerations in this paper on a recent experimental report Ref.~\cite{gity2018} where the overall composition of the oxide was found to be $\text{Bi}_2\text{O}_3$ and assume that the same stoichiometry is present at the interface.
In fact, the known bulk polymorphs can exist with different stoichiometries and have complicated structures.
A common oxide is $\alpha\text{-}\text{Bi}_2\text{O}_3$ which is stable at room temperature but has a monoclinic crystal structure with a basis consisting of $20$ atoms \cite{walsh2006}.
All the oxides do not allow easily for the construction of a suitable interface to the bismuth film.
More importantly, when making contact with the film, strong covalent bonds may destroy the bismuth surface bilayer as described previously for hydrogen bonding to the Bi surface.
As the crystal orientation is crucial regarding the film's electronic properties -- the bulk properties of bismuth are highly anisotropic -- we want to preserve the orientation of the unterminated sample.
Therefore, presumably, the bonding between the Bi\hkl(1 1 1) surface and the oxide may be similar to that between the bilayers of bulk bismuth.

\begin{figure}
\includegraphics{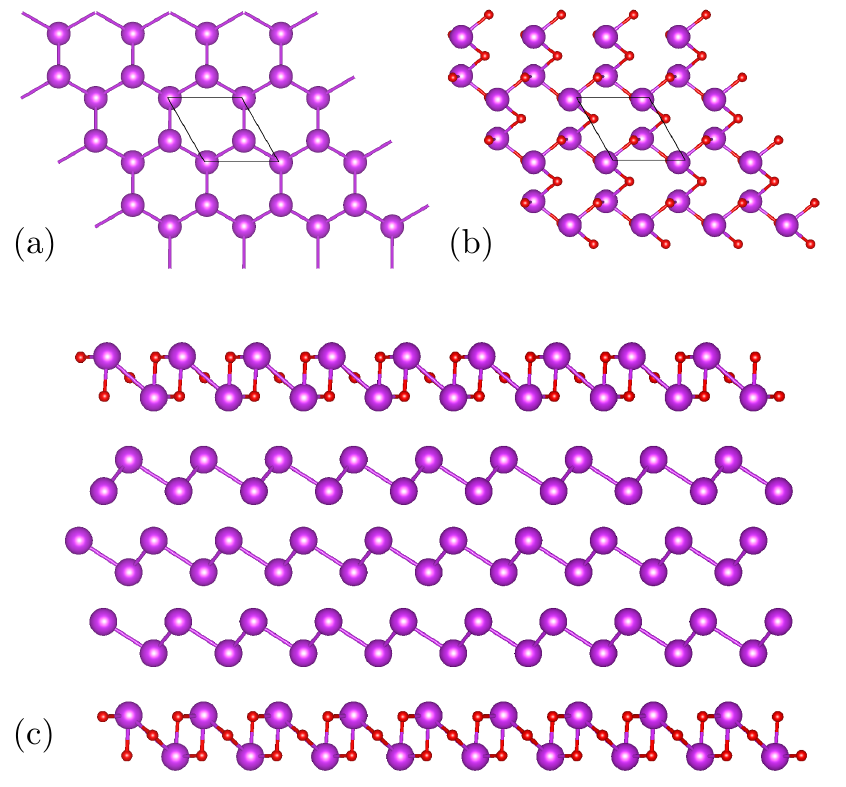}
\caption{ \label{fig:Bi2O3}
Construction of a model oxide structure on the Bi\hkl(1 1 1) surface.
The bismuth atoms in the bulk bilayer (a) all have three bonds.
These bonds are elongated in the oxide structure (b) by an intermediate oxygen atom per bond.
The hexagonal cell is drawn in black.
One O atom lies above its corresponding bismuth atom, another one almost vanishes below the next Bi, while a third O atom can be found in between the bismuth atoms.
The side view in (c) reveals that the oxide also has a layered structure and does not change the orientation of the underlying bismuth bilayers.}
\end{figure}

Based on the preceding qualitative discussion, we investigated a model structure for the native oxide derived from the crystal structure of bulk bismuth.
Note that a hexagonal crystal cell is employed again.
Only the surface bilayers are modified with additional oxygen atoms and the equilibrium structure is obtained by mi\-ni\-mi\-zing the total free energy.
Figure \ref{fig:Bi2O3} shows the structure in detail.
The bismuth atoms in the bilayer do not bond directly any more but the connection is made by an intermediate oxygen atom instead.
This effectively makes the bonds longer and allows for a considerable rearrangement of the atoms in the oxide layer.
Nevertheless, each bismuth atom has three bonds to neighboring atoms just as in the bulk.
The stoichiometry of our structure is $\text{Bi}_2\text{O}_3$.
From Fig.~\ref{fig:Bi2O3}, it can be seen clearly that the bilayered character of the bulk structure remains also in the oxide.

The absence of strong bonds implies a weak interaction between the film and the oxide.
We calculate the interfacial energy as a sum of two contributions
\begin{equation}
  E_{\text{interf}} = E_{\text{adh}} + E_{\text{rel}}~.
\end{equation}
Adhesion $E_{\text{adh}}$ leads to an attractive force between the three bismuth bilayers and the oxide layers.
The re\-la\-xa\-tion of the two material subsystems is taken into account by $E_{\text{rel}}$.
This separation allows us to determine each contribution based on a set of calculations with the same number of localized basis functions in order to account for basis set superposition errors.
For $E_{\text{adh}}$, the total free energies $E_{\text{bl}}$ and $E_{\text{ox}}$ of the subsystems were calculated by removing the atoms of the oxide layers (or bismuth bilayers, respectively) while keeping their basis functions in order to represent the wave functions (ghost atoms).
The energies $E^{\text{pre}}$ before and $E^{\text{post}}$ after the relaxation were determined without ghost atoms.
Note that we kept the lattice parameter fixed for all the calculations.
The final result for the interfacial energy is given by
\begin{eqnarray}
  \label{eq:Einterfacial}
  E_{\text{interf}} = E_{\text{bl+ox}}
  &-& \left[ E_{\text{bl}} + \left( E^{\text{post}}_{\text{bl}} - E^{\text{pre}}_{\text{bl}} \right) \right]\\*
  &-& \left[ E_{\text{ox}} + \left( E^{\text{post}}_{\text{ox}} - E^{\text{pre}}_{\text{ox}} \right) \right]\nonumber
  ~.
\end{eqnarray}
The interaction of the bilayers in a pure bismuth film with an equivalent thickness can be used as a reference point $E_{\text{ref}}$.
Therefore we also calculate the energies in Eq.~(\ref{eq:Einterfacial}) for a film with two additional bismuth bilayers instead of the oxide layers, i.e.~five bilayers in total.
Normalized to the surface area we find accordingly
\begin{eqnarray}
  E_{\text{interf}} &=& \SI{-0.29}{\electronvolt\per\nano\meter^2}~,\\
  E_{\text{ref}}    &=& \SI{-1.90}{\electronvolt\per\nano\meter^2}~.
\end{eqnarray}
\begin{figure*}
\includegraphics{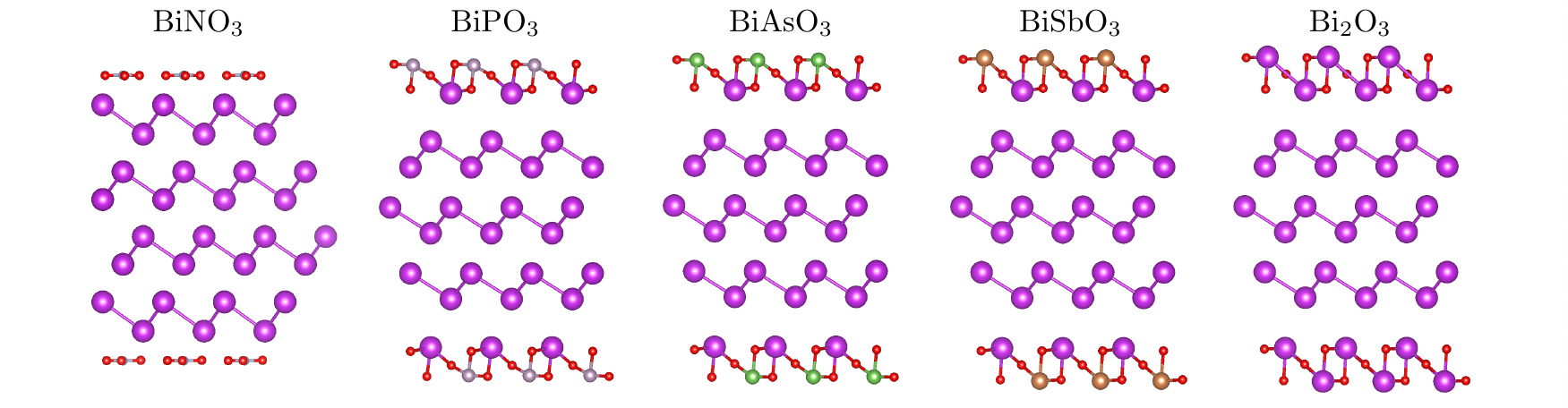}
\caption{ \label{fig:PAsSbOxideStruct}
Relaxed films with oxide termination including different group V elements.
All oxide layers except for $\text{Bi}\text{N}\text{O}_3$ have a very similar structure compared to $\text{Bi}_2\text{O}_3$.
The bond lengths within the oxide layer increase the heavier the group V element is.
The light nitrogen atoms detach with the bonded oxygen atoms so that the Bi-O distance becomes larger than \SI{2.5}{\angstrom}, which is an increase by more than \SI{10}{\percent} compared to the other structures.
Presumably, the N-O bonds are too short to allow for the formation of a comparable oxide.
}
\end{figure*}
The oxide layers interact much less with the underlying bismuth bilayer for this lattice constant than the weakly bonding bilayers do with each other.
Note that, taking the five bilayer film and an oxygen molecule in the triplet ground state as reference, we find that the oxide is stable with a formation energy of \SI{-29.0}{\electronvolt\per\nano\meter^2}.
Furthermore, this oxide model caps the film's bilayers and reduces the chemical interaction with the environment.

In order to generalize the oxide structure, we replaced the bismuth atoms in the oxide which are closest to the surface by other group V elements and obtained capping layers with similar structures.
The Bi atoms at the interface to the pure bismuth bilayers were not replaced to facilitate the bonding.
Alternative oxides can be par\-ti\-cu\-lar\-ly interesting to engineer nanoelectronic devices for example with improved electrostatic control of the gate over the channel region.
The corresponding structures for N, P, As, Sb, and Bi are shown in Fig.~\ref{fig:PAsSbOxideStruct}.
We find that the nitrogen structure is obviously very different to the other structures because each N bonds to three O atoms, thereby forming a flat nitrate-like molecule with a charge of $-1/2~\mathrm{e}$ and separates from the surface.
The bond length between N and O is \SI{1.26}{\angstrom} and the distance between O and Bi becomes larger than \SI{2.5}{\angstrom}, which is an increase by more than \SI{10}{\percent} compared to the other structures.
The remaining bismuth atoms are integrated into the underlying film and a fourth bilayer appears 
due to the introduction of two half bilayers at the top and bottom surfaces and a reorganization of the Bi bilayers during the relaxation of the structure.
We can conclude from the small bond length between nitrogen and oxygen that the formation of an oxide layer is not possible in this case.
In contrast to that, all other structures are very similar to the $\text{Bi}_2\text{O}_3$-terminated film discussed before.
Table \ref{tab:oxideSummary} gives a summary of some key properties of the oxide layers.
\begin{table}
\caption{ \label{tab:oxideSummary}
Properties of the oxide termination.
The interfacial energies of the oxide terminated structures are determined as discussed in the text.
Note that, as a reference, the interaction between pure bismuth bilayers is \SI{-1.90}{\electronvolt\per\nano\meter^2}.
By minimizing the total free energy we also find the strain in the oxide layers and their equilibrium lattice constants $a^{\text{ox}}_{\text{eq}}$ which are considerably larger than \SI{4.473}{\angstrom}.
}
\begin{ruledtabular}
\bgroup 
\def\arraystretch{1.5}
\begin{tabular}{lcccc}
                & $\text{Bi}\text{P}\text{O}_3$
                & $\text{Bi}\text{As}\text{O}_3$
                & $\text{Bi}\text{Sb}\text{O}_3$
                & $\text{Bi}_2\text{O}_3$ \\
\hline
$E_{\text{interf}} \left[\frac{\si{\electronvolt}}{\si{\nano\meter^2}}\right]$ 
                            & $-0.59$   
                            & $-0.28$   
                            & $-0.26$   
                            & $-0.29$ \\
$\text{strain} \left[\si{\percent}\right]$ & $-9.8$ & $-10.7$ & $-13.0$ & $-12.1$ \\
$a^{\text{ox}}_{\text{eq}} \left[\si{\angstrom}\right]$ & $4.96$ &  $5.01$ &  $5.14$ &  $5.09$ \\
\end{tabular}
\egroup 
\end{ruledtabular}
\caption{ \label{tab:oxideSummaryAtEqA}
Properties of the oxide termination at the equi\-li\-bri\-um lattice parameter of bulk bismuth, i.e.~$a=\SI{4.532}{\angstrom}$.
In this case, the equivalent reference is \SI{-2.08}{\electronvolt\per\nano\meter^2}, i.e.~the interaction between pure bismuth bilayers is slightly higher than for the constrained lattice parameter with \SI{4.473}{\angstrom}.
}
\begin{ruledtabular}
\bgroup 
\def\arraystretch{1.5}
\begin{tabular}{lcccc}
                & $\text{Bi}\text{P}\text{O}_3$
                & $\text{Bi}\text{As}\text{O}_3$
                & $\text{Bi}\text{Sb}\text{O}_3$
                & $\text{Bi}_2\text{O}_3$ \\
\hline
$E_{\text{interf}} \left[\frac{\si{\electronvolt}}{\si{\nano\meter^2}}\right]$
                            & $-0.60$
                            & $-0.31$
                            & $-0.28$
                            & $-0.30$ \\
$\text{strain} \left[\si{\percent}\right]$ & $-8.6$ & $-9.5$ & $-11.8$ & $-11.0$ \\
\end{tabular}
\egroup 
\end{ruledtabular}
\end{table}
We find that the interfacial energy for each oxide is smaller by at least a factor of three compared to the reference energy of the unterminated five bilayer film.
Interestingly however, $|E_{\text{interf}}|$ has a maximum for $\text{Bi}\text{P}\text{O}_3$ and decreases for the other oxides which coincides with an increase in strain within the oxide layer.
For com\-pa\-ri\-son, Table \ref{tab:oxideSummaryAtEqA} summarizes the same properties for structures with the lattice dimensions of bulk bismuth which is applicable, e.g., at the surfaces of thick bismuth films where the influence of the substrate is negligible.
Here the stress is reduced but the interaction between the oxides and the film is still smaller than for the reference bismuth system which can be due to the persisting lattice mismatch with the bismuth film.

The bismuth atoms within the oxides close to the interface can be expected to facilitate the bonding to the bilayers because they remain approximately in their bulk position.
The corresponding bond lengths are summarized in Table \ref{tab:oxideBondLengthsToFilm}.
In agreement with the small interfacial energy, the distance between the remaining bismuth atoms in the oxide and the Bi bilayers is increased compared to the equivalent bonds in a five-bilayer thick bismuth film without termination which only have a length of \SI{3.67}{\angstrom}.
Nevertheless, these changes are small enough to not significantly change the equilibrium distance between the bismuth bilayers and the oxide layers compared to that between two equivalent bismuth bilayers.

\begin{table}
\caption{ \label{tab:oxideBondLengthsToFilm}
Distance between the bismuth atoms in the oxide and their next neighbors in the adjacent bismuth bilayers in angstroms.
In a five-bilayer bismuth film, the equivalent three bonds all have the same shorter bond length of $\SI{3.60}{\angstrom}$.}
\begin{ruledtabular}
\bgroup 
\def\arraystretch{1.5}
\begin{tabular}{lcccc}
                & $\text{Bi}\text{P}\text{O}_3$
                & $\text{Bi}\text{As}\text{O}_3$
                & $\text{Bi}\text{Sb}\text{O}_3$
                & $\text{Bi}_2\text{O}_3$ \\
\hline
 bond length Bi-Bi & $3.97$ & $4.05$ & $4.08$ & $4.07$ \\
                   & $4.01$ & $4.07$ & $4.09$ & $4.08$ \\
                   & $4.03$ & $4.07$ & $4.10$ & $4.08$ \\
\end{tabular}
\egroup 
\end{ruledtabular}
\caption{ \label{tab:oxideBondLengths}
Bond lengths within the oxide layers in angstroms.
X denotes the group V element which is used to replace the bismuth atoms at the surface of the terminated film.
In contrast to the X-O bonds, the length of the Bi-O bonds stays almost constant within and among the different structures.
}
\begin{ruledtabular}
\bgroup 
\def\arraystretch{1.5}
\begin{tabular}{lcccc}
                & $\text{X}=\text{P}$
                & $\text{X}=\text{As}$
                & $\text{X}=\text{Sb}$
                & $\text{X}=\text{Bi}$ \\
                & $\text{Bi}\text{P}\text{O}_3$
                & $\text{Bi}\text{As}\text{O}_3$
                & $\text{Bi}\text{Sb}\text{O}_3$
                & $\text{Bi}_2\text{O}_3$ \\
\hline
 bond length X-O  & $1.59$ & $1.80$ & $2.03$ & $2.18$ \\
                  & $1.65$ & $1.83$ & $2.04$ & $2.18$ \\
                  & $1.74$ & $1.91$ & $2.07$ & $2.19$ \\
 bond length Bi-O & $2.16$ & $2.14$ & $2.14$ & $2.17$ \\
                  & $2.16$ & $2.16$ & $2.16$ & $2.18$ \\
                  & $2.18$ & $2.17$ & $2.19$ & $2.20$ \\
\end{tabular}
\egroup
\end{ruledtabular}
\end{table}

Finally, Table \ref{tab:oxideBondLengths} shows all the bond lengths within the oxide.
The bonds to the oxygen atoms are shorter for lighter group V atoms like phosphorus -- denoted by X in the table -- and increase in length for the heavier elements.
Because of the fixed lattice constant, the bond angles change accordingly.
In contrast to that, the Bi-O bonds and angles at the interface to the bilayers do not change considerably throughout the structures.

\section{Discussion}
In the first half of this paper, we showed how the structure of thin Bi\hkl(1 1 1) films changes if they are terminated with hydrogen or hydroxyl.
This type of termination is not thermally stable under the assumptions made for equilibrium conditions. However, hydrogen was considered as a simple model system to investigate the chemical interaction with reactive gases and can also provide insight into the interaction with the dangling bonds of the silicon substrate and the formation of the wetting layer.

We discussed different hydrogen concentrations starting with one single hydrogen per surface atom.
The DFT calculations show that
hydrogen reacts with the surface and has a profound effect on the structure close to the surface which is especially important for thin films where the surface bonding induces a reorientation throughout the film from Bi\hkl(1 1 1) to Bi\hkl(1 0 0).
A continuous transition of the bond lengths with increasing thickness can be observed in the middle of the film which ultimately leads to an inner region which is not affected by the reorientation.
Specifically, in films with a thickness of more than five bilayers, the first two bilayers from the surface are clearly affected by the hydrogen termination while the middle layers remain in the Bi\hkl(1 1 1) orientation.
Our results are supported by bulk energy calculations which show that the substrate induced strain stabilizes the Bi\hkl(1 1 1) orientation in the middle of the thick films with respect to Bi\hkl(1 0 0).
However, the exact number of bilayers which is affected by the surface termination may depend on the accuracy of the calculation as the bond lengths transform continuously from the surface into the bulk and the distinction between regimes of different orientations is based on nearest neighbors.
There may also be an energy barrier which has to be overcome to change the orientation of the crystal.
The structure of the surface layers is preserved even in thick films which have a Bi\hkl(1 1 1) orientation in the majority of the film.

The same type of reorganization is also observed for other monovalent passivants like hydroxyl.
This implies that the effect is not unique to hydrogen but is a more general feature.
Increasing the hydrogen surface density to two hydrogen atoms per surface atom has basically the same effect on the structure.

We emphasize, that in contrast to the thin film allotropes discussed in the introduction, we observe a reorientation which is driven by the chemical interaction at the surface and not motivated by a tradeoff of bulk and surface energy due to unsaturated dangling bonds.

Furthermore, we showed that the orientation of the bilayers in thin Bi\hkl(1 1 1) films can be preserved for an arbitrary film thickness if the bismuth atoms form three bonds to the passivating species, e.g.,~in the case of an even higher hydrogen surface density. We found that in this case charge-neutral bismuthine molecules ($\text{Bi}\text{H}_3$) form on the surface of the film.
With the exception of Refs.~\cite{nagao2004,saito2004}, we are not aware that such a passivation scheme was taken into account previously.
The authors of \cite{nagao2004} find no major influence of their $\text{Bi}_2\text{H}_2$ and $\text{Bi}\text{H}_3$ wetting layer on the relaxation of their structures which is in agreement with our observation that $\text{Bi}\text{H}_3$ does not form a bond to the surface.
We want to stress however, that bismuthine is not stable \cite{amberger1961} but highly reactive and thus can not be considered to be a realistic surface termination.

In the second part of this paper, we presented a simplified model for the native oxide on the Bi\hkl(1 1 1) surface.
This model offers several advantages over the rather artificial surface passivation which was discussed before.
First of all, it is a better approximation to the chemistry of the native oxide as each Bi atom forms three bonds to the surrounding O atoms and thus the stoichiometry is the same as in bulk $\text{Bi}_2\text{O}_3$.
Furthermore, the bilayer orientation in the film is not perturbed by the passivation which is achieved by a weak interaction between the film and the oxide.
Within the scope of our model the oxide forms a layer similar to the bilayers of the bulk system and helps to protect the surface from further chemical interaction.
We emphasize that the oxide layers are stable. Thus, by direct comparison, the metastability of the covalently bonded terminations further supports the plausibility of our oxide model and the weak bonding to the surface.
To the best of our knowledge, no comparable model exists in the literature although the surface plays a considerable role in the properties of thin films or nanowires.

Finally, more complex models for oxide layers including other group V elements were introduced.
We found the same layered structure which preserves the orientation of the underlying bilayers.
The compressive stress in the oxide layer is reduced once the films are thick enough to re-establish the bulk lattice constant.
Note that the chemical modification with phosphorus can apparently increase the interaction with the bismuth bilayers and lead to stronger adhesion.
Future work will investigate the consequences for the electronic properties of the system with more predictive methods for the electronic structure than the GGA.
A key topic for investigation is the interaction of the oxide layers with the metallic surface states known to exist on the Bi\hkl(1 1 1) surface.

We highlight that the alterations to the structure have a direct impact on the electronic properties of the system as these depend strongly on the crystal orientation.
Thus the surface will have a crucial influence on device operation.
This is especially true regarding the long-standing discussion about the existence of the quantum confinement effect in bismuth.
A good introduction to the controversies regarding the semiconducting or metallic character of very thin films is given in Ref.~\cite{zhu2016}.
Recent investigations suggest that the surface states could effectively prevent the nanosystems from becoming semiconducting and that a distinction between them and the bulk states becomes difficult for very thin films \cite{xiao2012,hirahara2007,aitani2014,hirahara2015,zhu2016,hirahara2018,zhu2018,kroeger2018}.
A better understanding of the surface structure and its impact on the electronic structure will therefore be crucial in order to enable bismuth nanoelectronic or spintronic devices.

\begin{acknowledgments}
\section{Acknowledgments}
This work has been funded by Science Foundation Ireland through the Principal Investigator Award No.~13/IA/1956.
Atomistic structures were visualized with the \textsc{vesta} software \cite{momma2011}.
The authors want to thank Lida Ansari and Alfonso Sanchez-Soares for helpful discussions.
\end{acknowledgments}


\bibliography{./bibliography}


\end{document}



\title{\textit{Supplemental Material:} Structural modification of thin Bi\hkl(1 1 1) films by passivation and native oxide model}


\author{Christian \surname{König}}
\affiliation{Tyndall National Institute, University College Cork, Lee Maltings, Cork T12 R5CP, Ireland}
\author{Stephen \surname{Fahy}}
\affiliation{Tyndall National Institute, University College Cork, Lee Maltings, Cork T12 R5CP, Ireland}
\affiliation{Department of Physics, University College Cork, College Road, Cork T12 K8AF, Ireland}
\author{James C. \surname{Greer}}
\affiliation{Nottingham Ningbo New Materials Institute and Department of Electrical and Electronic Engineering, University of Nottingham Ningbo China, 199 Taikang East Road, Ningbo, 315100, China}

\date{\today}


\maketitle

\section{Stability of the hydrogen and hydroxyl terminated surfaces}
\begin{figure*}
  \begin{overpic}{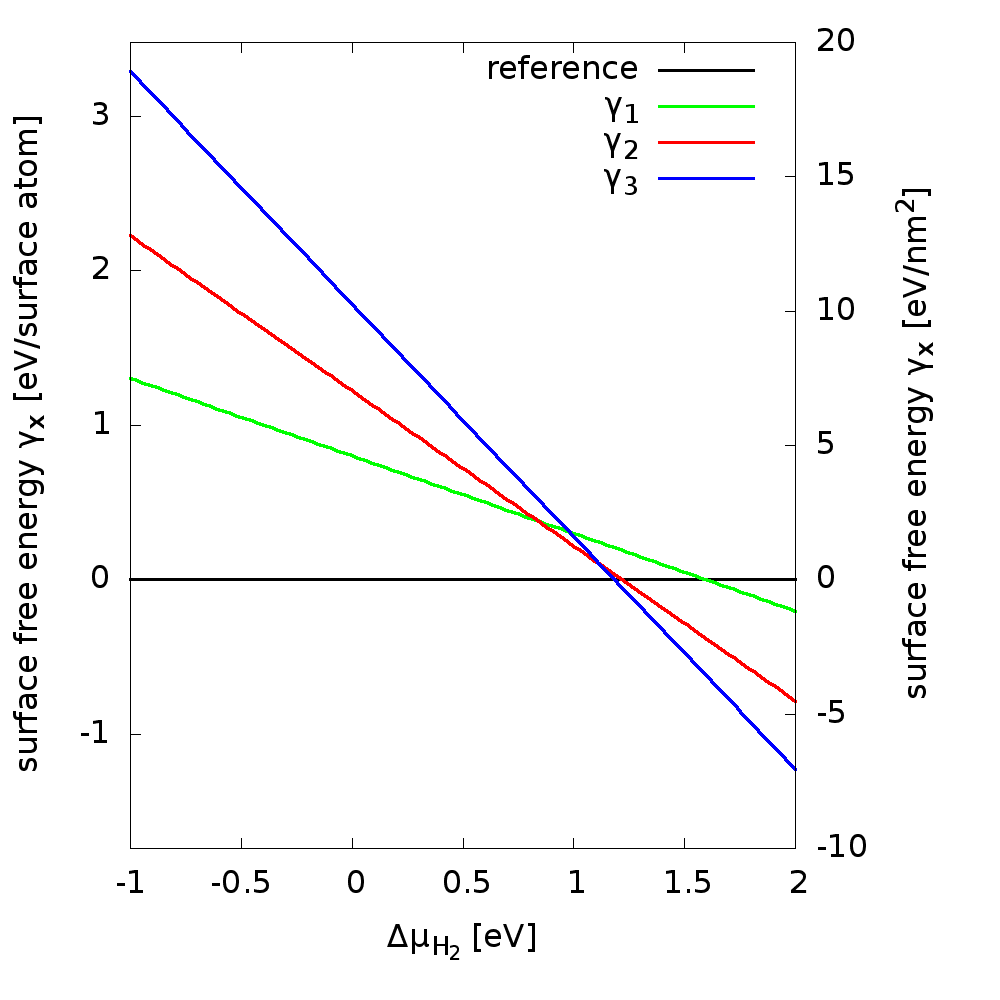}
    \put(5,5){(a)}
  \end{overpic}
  \begin{overpic}{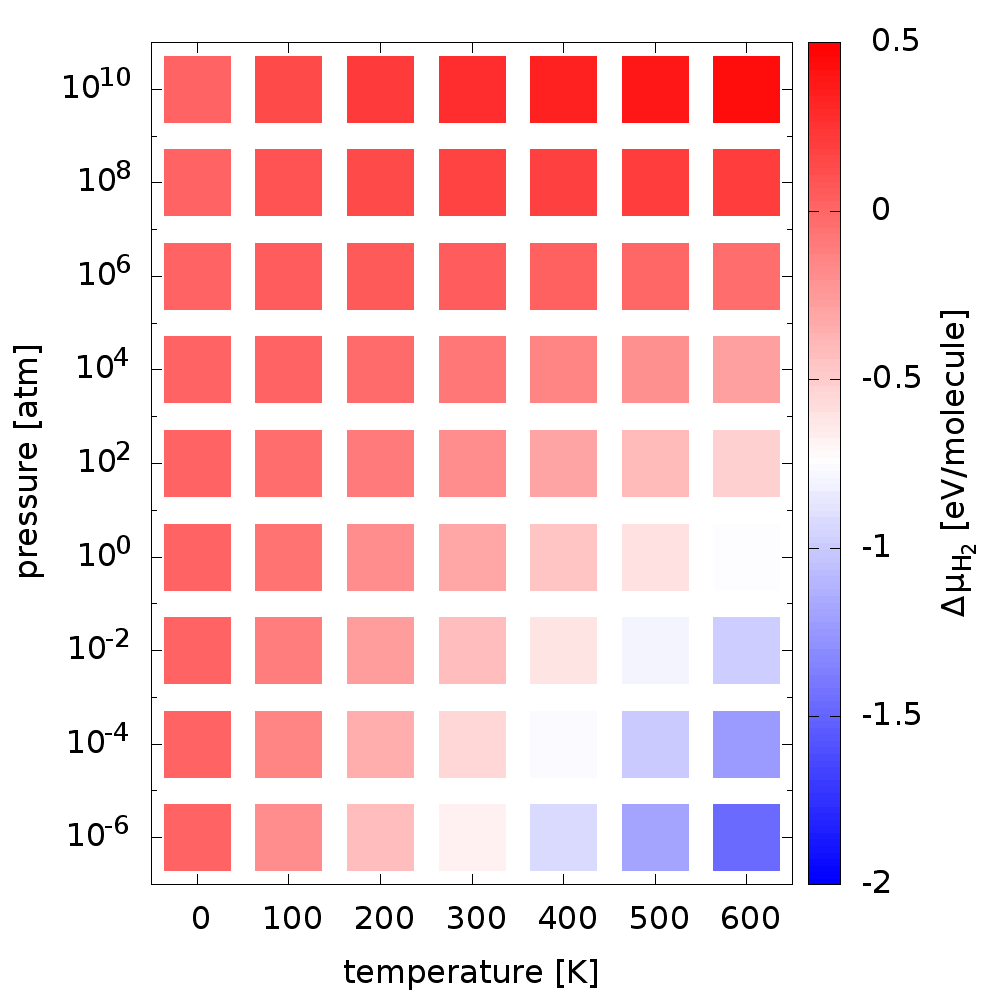}
    \put(5,5){(b)}
  \end{overpic}
\caption{ \label{fig:H}
          (a) Stability of the surfaces terminated with $x$ hydrogen atoms per surface atom as a function of changes of the chemical potential of the hydrogen gas with respect to \SI{0}{\kelvin}. The error introduced by basis set superposition is negligible (approximately \SI{1}{\percent}). (b) Dependence of $\Delta\mu_{\text{H}_2}$ on pressure and temperature. We find that the terminated surfaces are only stable if the chemical potential of the hydrogen gas can be substantially increased. This corresponds to a very high gas pressure exceeding typical experimental conditions.
        }
\end{figure*}

The stability of the terminated surfaces depends on the chemical potentials $\mu_i$ of the involved reactants $i$ which can be controlled in the experiment by adjusting the temperature and the gas pressure (e.g.~of H$_2$). We employ a formalism similar to Refs.~\cite{reuter2001,reuter2003} where the Gibbs free energy $G$ is approximated by the DFT total energy and vibrational contributions are neglected for the terminated slab. The formation energy is then given by
\begin{eqnarray}
  E^{\text{formation}} &=& G - \sum_i N_i \mu_i~,\\
                       &\approx& E^{\text{DFT}} + pV - \sum_i N_i \mu_i~.
\end{eqnarray}
with the particle numbers $N_i$.
The surface free energy $\gamma$ is then the difference between the formation energies of the terminated system and the unterminated slab divided by the surface area $A$. Both systems need to have the same number of particles which can be achieved by additional gas molecules in the vacuum. Let $\gamma_x$ be the surface free energy of a system terminated with $x$ hydrogen atoms per surface atom. The orthorhombic slab has two surface atoms on each side, therefore
\begin{equation}
  \gamma_x\left(T,p_{\text{H}_2}\right) = \left[ E^{\text{DFT}}_x - E^{\text{DFT}}_{x=0} - 4x\mu_\text{H}\left(T,p_{\text{H}_2}\right) \right]/A ~.
\end{equation}
By assuming that the chemical potential of a single H atom at \SI{0}{\kelvin} is half of the DFT energy of a H$_2$ molecule, we can make the ansatz
\begin{eqnarray}
  \mu_{\text{H}}=\frac{1}{2}\left( E^{\text{DFT}}_{\text{H}_2} + \Delta \mu_{\text{H}_2}\left(T,p_{\text{H}_2}\right) \right) ~,\\
  \text{where~~}\Delta \mu_{\text{H}_2}\left(\SI{0}{\kelvin},p_{\text{H}_2}\right)=0~.~~~~~
\end{eqnarray}
The chemical potential can be interpreted as a parameter in the theoretical considerations and the results are shown in Fig.~\ref{fig:H}(a).
Apparently, the terminations are not stable at \SI{0}{\kelvin} corresponding to $\Delta \mu_{\text{H}_2}=0$ and the chemical potential has to increase by more than \SI{1}{\electronvolt} to obtain a stable structure. However, in the DFT calculation the hydrogen atoms on the surface are too far apart to react with each other during the atomic relaxation. The bonds to the film show that the termination is stable with respect to single hydrogen atoms.

Following the reasoning in Refs.~\cite{reuter2001,reuter2003}, assuming an ideal gas it can be shown by calculating the total derivative of $\mu$ that with respect to \SI{0}{\kelvin} and an arbitrary pressure $p_0$
\begin{eqnarray}
  \Delta \mu(T,p) = \frac{H-ST}{N}\bigg|_{0K,p_0}^{T,p_0} + kT\ln\left(\frac{p}{p_0}\right)~.
\end{eqnarray}
Note that $\Delta \mu=0$ for \SI{0}{\kelvin} independent of the pressure which is why $p_0$ is arbitrary.
Experimental values for the enthalpy $H$ and entropy $S$ can be found in the literature \cite{stull1971} for atmospheric pressure (\SI{1}{atm} or \SI{1013.25}{\hecto\pascal}). Fig.~\ref{fig:H}(b) shows how the chemical potential of H$_2$ changes with temperature and gas pressure. We find that the changes of $\mu_{\text{H}_2}$ which would be necessary to stabilize the surface are only accessible with very high pressure.

\begin{figure*}
  \begin{overpic}{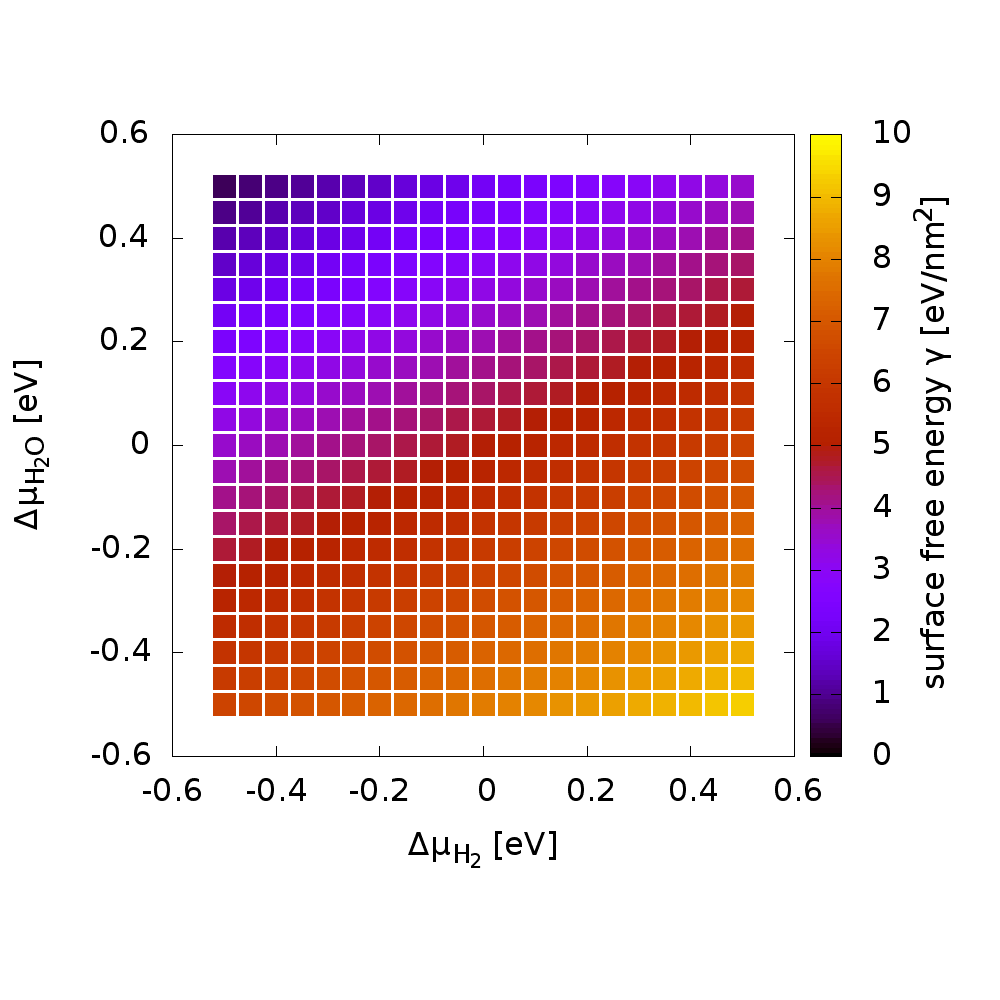}
    \put(5,5){(a)}
  \end{overpic}
  \begin{overpic}{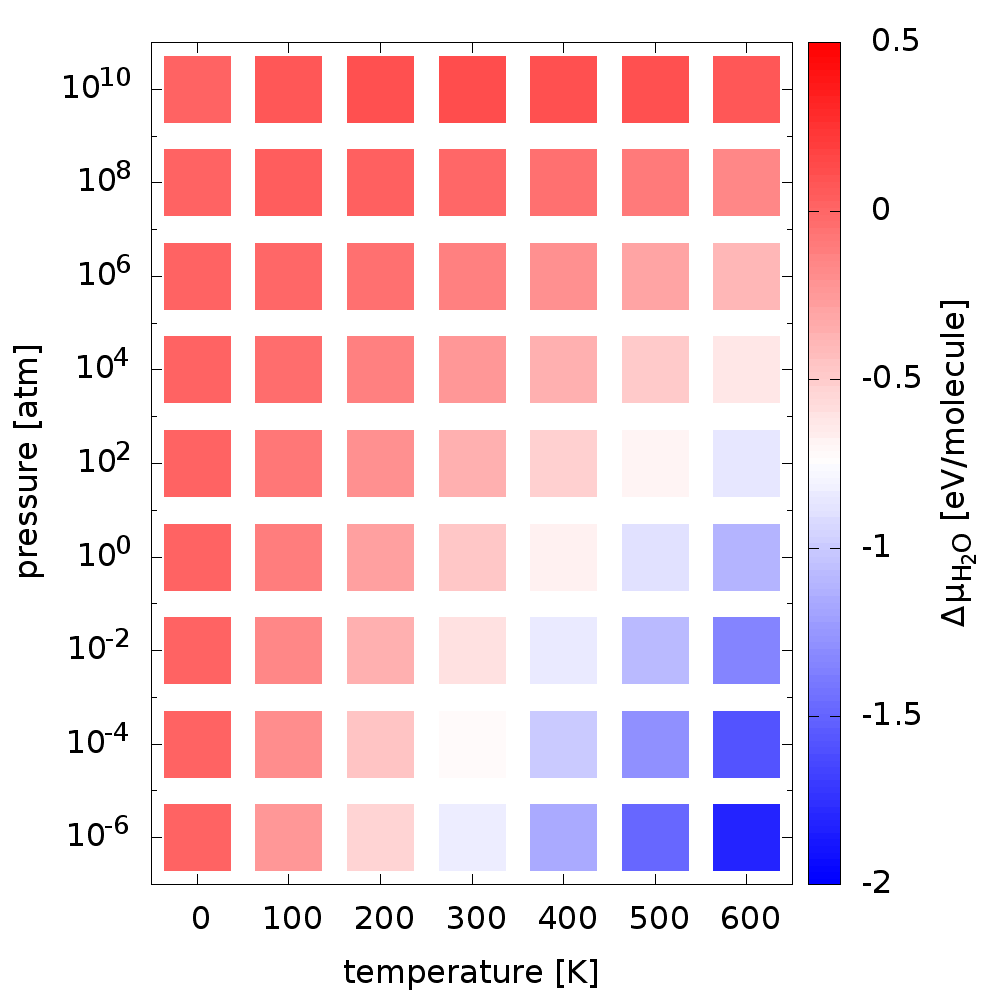}
    \put(5,5){(b)}
  \end{overpic}
\caption{ \label{fig:OH}
          (a) Stability of the OH-terminated surface (one OH per surface atom) as a function of changes of $\mu_{\text{H}_2}$ and $\mu_{\text{H}_2\text{O}}$ with respect to \SI{0}{\kelvin}. Water molecules are assumed to dissociate at the surface, leading to a passivated surface and the formation of H$_2$. At $\Delta\mu_{\text{H}_2}=\Delta\mu_{\text{H}_2\text{O}}=0$ we find $\gamma=\SI{4.95}{\electronvolt\per\nano\meter^2}$. (b) Dependence of $\Delta\mu_{\text{H}_2\text{O}}$ on pressure and temperature. The terminated surface is not stable under typical experimental conditions as we have already observed for the hydrogen terminated surfaces.
          }
\end{figure*}

The same considerations can be applied to the hydroxyl termination. We assume that water reacts with the surface of the bismuth films which leads to the formation of H$_2$. Thus, there are two chemical potentials $\mu_{\text{H}_2}$ and $\mu_{\text{H}_2\text{O}}$ which have to be taken into account. In line with the paper we restrict our discussion here to a single OH per surface atom. Fig.~\ref{fig:OH}(a) shows the surface free energy as a function of the chemical potentials. The influence of the experimental control parameters on $\mu_{\text{H}_2\text{O}}$ are displayed in Fig.~\ref{fig:OH}(b). Again, under the assumptions made, we find that the structure is not stable under typical experimental conditions.
We stress however that the terminations above should rather be considered as model systems with a specified amount of covalent bonds to the surface.

\bibliography{./bibliography}

